\input harvmac
\writedefs
\sequentialequations

\def\comment#1{}

\def\df#1#2{{\displaystyle{#1 \over #2}}}

\def\+{^\dagger}

\def\i{i}

\def \chi {\chi}\def\r {\rho}

\def \inv {^{-1}}
\def \ov {\over }

\def\np {{  Nucl. Phys. }}
\def \pl {{  Phys. Lett. }}
\def \mpl {{ Mod. Phys. Lett. }}
\def \prl {{  Phys. Rev. Lett. }}
\def \pr  {{ Phys. Rev. }}


\Title{
 \vbox{\baselineskip10pt
  \hbox{PUPT-1679}
  \hbox{MIT-CTP-2610}
  \hbox{hep-th/9701187}
 }
}
{
 \vbox{
  \centerline{ Black Hole Greybody Factors and}
  \vskip 0.1 truein
  \centerline{Absorption of Scalars by Effective Strings }
 }
}
\vskip -25 true pt

\centerline{
 Igor R. Klebanov and 
Samir D.  Mathur\foot{On leave from MIT, Cambridge, MA 02139.} }
\centerline{\it Joseph Henry Laboratories, 
Princeton University, Princeton, NJ  08544}

\bigskip
\centerline {\bf Abstract}
\bigskip
\baselineskip10pt
\noindent

We compute the greybody  factors for classical black holes in a domain
where two kinds of charges and their anticharges are excited by
the extra energy over extremality. We compare the result to the greybody
factors expected from an effective string model which
was earlier shown to give the
correct entropy. In the regime where the left and right moving
temperatures are much smaller than the square root of the
effective string tension, we find a non-trivial greybody factor
which agrees with the effective string model. However, if the
temperatures are comparable with the square root of 
the effective string tension,
 the greybody factors agree only at the
leading order in energy. Nevertheless,
there are several interesting relations
between the two results, suggesting that a modification
of the effective string model might lead to better agreement.

\Date {January 1997}

\noblackbox
\baselineskip 14pt plus 1pt minus 1pt


\lref\mast{J.M.~Maldacena and A.~Strominger, Rutgers preprint RU-96-78, 
hep-th/9609026.}
\lref\mst{J.M.~Maldacena and A.~Strominger, \prl 77 (1996) 428, 
hep-th/9603060.}

\lref\HP{G. Horowitz and J. Polchinski, hep-th/9612146.}

\lref\kr{B.~Kol and A.~Rajaraman, Stanford preprint SU-ITP-96-38, 
SLAC-PUB-7262, hep-th/9608126.}

\lref\dvv{R. Dijkgraaf, E. Verlinde and H. Verlinde, hep-th/9603126;
hep-th/9604055.}

\lref\gkk{G.~Gibbons, R.~Kallosh and B.~Kol, Stanford preprint 
SU-ITP-96-35, hep-th/9607108.}

\lref\fkk{S. Ferrara and  R. Kallosh, hep-th/9602136; hep-th/9603090;
S. Ferrara, R. Kallosh, A. Strominger, Phys. Rev. D{52} (1995)
5412, hep-th/9508072. }

\lref\hmf{{\it Handbook of Mathematical Functions}, 
M.~Abramowitz and I.A.~Stegun, eds. (US Government Printing Office,
Washington, DC, 1964) 538ff.}

\lref\GR{I.S.~Gradshteyn and I.M.~Ryzhik, {\it Table of Integrals, 
Series, and Products}, Fifth Edition, A.~Jeffrey, ed. (Academic Press:
San Diego, 1994).}

\lref\sv{A.~Strominger and C.~Vafa, \pl B379 (1996) 99, hep-th/9601029.}

\lref\cm{C.G.~Callan and J.M.~Maldacena, \np B472 (1996) 591, hep-th/9602043.}

\lref\ms{J.M.~Maldacena and L.~Susskind, Stanford preprint
SU-ITP-96-12, hep-th/9604042.}

\lref\dmw{A.~Dhar, G.~Mandal and S.~R.~Wadia, 
\pl B388 (1996) 51, hep-th/9605234.}

\lref\dm{S.R.~Das and S.D.~Mathur,
Nucl. Phys. {\bf B478} (1996) 561, hep-th/9606185. 
 }

\lref\dmI{S.R.~Das and S.D.~Mathur,  hep-th/9601152.}

\lref\hms{G.~Horowitz, J.~Maldacena and A.~Strominger, \pl B383 (1996) 151,
hep-th/9603109.}\comment{Journal reference added}

\lref\HS{G.~Horowitz  and A.~Strominger, \prl  77 (1996) 2368, 
hep-th/9602051.}

\lref\GK{S.S.~Gubser and I.R.~Klebanov, 
Nucl. Phys. B482 (1996) 173, hep-th/9608108.}

\lref\GKtwo{S.S.~Gubser and I.R.~Klebanov, Phys. Rev. Lett. 77
(1996) 4491, hep-th/9609076. }

\lref\CGKT{
C.G.~Callan, Jr., S.S.~Gubser, I.R.~Klebanov and A.A.~Tseytlin,
hep-th/9610172.}

\lref\dgm{S.~Das, G.~Gibbons and S.~Mathur, hep-th/9609052.}

\lref\GKP{S.S.~Gubser, I.R.~Klebanov and A.W.~Peet, \pr D54 (1996) 3915,
hep-th/9602135. }

\lref\ktI{I.R.~Klebanov and A.A.~Tseytlin, 
Nucl. Phys. B479 (1996) 319, hep-th/9607107.}

\lref\KT{I.R. Klebanov and A.A. Tseytlin, \np B475 (1996) 179, 
hep-th/9604166.}

\lref\kt{I.R. Klebanov and A.A. Tseytlin, \np B475 (1996) 165, 
hep-th/9604089.}

\lref\HK{A. Hanany and I.R. Klebanov, \np B482 (1996) 105,
hep-th/9606136.}

\lref\KK{I.R. Klebanov and M. Krasnitz, hep-th/9612051.}

\lref\at{ A.A. Tseytlin, \np B475 (1996) 49, hep-th/9604035.}

\lref\ATT{A.A. Tseytlin, \mpl A11 (1996) 689,  hep-th/9601177.}

\lref\CY{M. Cveti\v c and D. Youm, \pr D53 (1996) 584, hep-th/9507090;
Contribution to `Strings 95',  hep-th/9508058. }

\lref\CT{M. Cveti\v c and  A.A.  Tseytlin, \np B478 (1996) 181,
hep-th/9606033.}\comment{Journal reference added}

\lref\CTT{M. Cveti\v c and  A.A.  Tseytlin, \pr D53 (1996) 5619, 
 hep-th/9512031.}

\lref\CYY{M. Cveti\v c and D. Youm, hep-th/9603100.} 

\lref\maldacena{J. Maldacena, hep-th/9605016.}

\lref\myers{C. Johnson, R. Khuri and R. Myers, 
Phys. Lett. B378 (1996) 78, hep-th/9603061.} 

\lref\HS{G.T. Horowitz  and A. Strominger, \prl 77 (1996) 2368,
hep-th/9602051.}

\lref\LWW {F. Larsen and F. Wilczek,  \pl B375 (1996) 37,
hep-th/9511064; hep-th/9609084.}

\lref\dmI{S.R.~Das and S.D.~Mathur, Phys. Lett. B375 (1996) 103, 
 hep-th/9601152.}

\lref\hrs{E.~Halyo, B.~Kol, A.~Rajaraman and L.~Susskind, hep-th/9609075;
E.~Halyo, hep-th/9610068.}

\newsec{Introduction}

In recent times there has been considerable progress in understanding
the properties of black holes from a microscopic fundamental
theory of gravity -- string theory. The state content of nonperturbative
string theory appears to be just correct to give the entropy
of extremal black holes, when we estimate the number of states of
string theory carrying the same charges as the black hole \sv.
This result on the entropy extends to a variety of situations where a 
black hole is close to extremality 
\refs{\cm,\HS,\maldacena}. The entropy of certain near-extremal
branes sometimes agrees with the Bekenstein-Hawking entropy up to
factors of order unity \refs{\GKP,\kt} whose origin was recently
explained in \HP.
These developments follow earlier suggestions
of Russo and Susskind \ref\rususs{J. Russo and L. Susskind,
\np B437 (1995) 611.} that the large number of
states of a string with given mass might be related to
the large entropy of black holes with the same mass, and calculations of
Sen \ref\sen{A. Sen, \np B440 (1995) 421
and \mpl A10 (1995) 2081. } which showed such agreement 
up to a numerical coefficient for elementary 
string states corresponding to extreme black holes.

Somewhat surprisingly, the rates of exciting and de-exciting the
states of string theory that correspond to a black hole agree,
at low energies,
with the rates of absorption and Hawking radiation 
 for the black hole. The string calculation is carried out at
weak coupling, while the black hole with a classical
horizon is a good description for strong coupling. There is no
a priori reason for this agreement at different values of the coupling;
nevertheless it appears to work in a remarkably
large class of low energy interactions. The leading term
for minimally coupled scalars in 5 dimensions
was computed in \dm, for the string state and for the classical
 black hole, and the above mentioned agreement was found.
(It had been shown in \cm, \dmw\ that the two rates
should agree up to a constant of proportionality.)
The   agreement at leading  order for charged scalars in 4 and
 5 dimensions was demonstrated
in \GK. In \mast{} such calculations were
extended (in 5 dimensions)
to the case where the emitted energy is comparable to the 
appropriately defined left and right moving temperatures,
which yields agreement for the
greybody factor of the black hole in this domain. Such greybody
factors agree also in 4 dimensions \GKtwo, and 
for the absorption of the non-minimally coupled `fixed scalars' 
\refs{\fkk,\gkk,\kr}
in 5 dimensions \CGKT\ and in 4 dimensions \KK .

In the above calculations the model used for the effective 
description of the stringy black holes
was that of a single long string that
absorbs and emits gravitons by coupling them to 
its vibration modes.\foot{In \ref\dasmathurthree{
S. Das and S.D. Mathur, hep-th/9607149} it
was shown that the elementary string quantized by the 
Polyakov prescription leads to the same cross sections at leading order in 
energy.} These modes are assumed
to be the lowest energy excitations in the system, which implies that
the moduli of the spacetime are chosen to suppress the excitations
of other kinds of charges and anti-charges. What happens when the
moduli are such that the excitation of one other kind of charge pair
becomes relevant? 

The simplest case of a second excitation which becomes relevant is that
the string which carried the vibrations (the `momentum' and `antimomentum'
 modes) can also  be excited to give additional pairs of `strings-antistrings'.
But the creation of a string-antistring pair can be thought of as
a segment of the original string that bends over to run for a while
in the opposite direction in target space, after which it runs in the
forward direction again. Thus, such `bends' are equivalent to
the string-antistring pairs.

When we quantize the  vibrations of the D-string as a 1-dimensional gas
of massless particles living on the D-string, then we are in the
limit where these vibrations are small amplitude,
 long wavelength  transverse displacements of the  string. Thus the `bends'
that might correspond to the string-antistring excitations are not included.
But if we quantize the string by the Polyakov prescription, then all
configurations of the string are included, and so we expect that
both the momentum-antimomentum and the string-antistring
excitations are taken into account.

Indeed it was shown in \maldacena\ that using
a Polyakov quantized effective string the entropy of the
5 dimensional black hole was correctly reproduced in the domain where
one of the charges (given by the 5-D-branes) is large, 
but the momentum, antimomentum, string winding and string antiwinding
excitations are all comparable.\foot{A description of extremal black holes
using a similar non-critical string theory was worked out in
\dvv.} The tension of this effective string
was found to be equal to that of a D-string divided by the number of 
the 5-branes.\foot{A  rescaling of the string tension
was also advocated for the NS-NS charged black holes in
\refs{\LWW,\CTT}.}
In \hrs\ it was shown that if we
assume that a Polyakov quantized string has an effective central charge of 6,
then its low energy absorption cross section for
minimal scalars equals the area of the horizon of the corresponding 
black hole, which is classically the correct cross section
for the black hole \dgm. This leading order in energy calculation
was carried out in more detail and extended to the 4-dimensional case in 
\ref\halyo{E. Halyo, hep-th/9610068, hep-th/9611175. }.

In this paper we investigate the absorption and emission of minimally
coupled scalars from a black hole
which has two charges and the nonextremality parameter all comparable
to each other. This means that we can excite two kinds of branes and
antibranes, and thus this case cannot be covered by a model where
we consider small transverse vibrations of an effective string. For
the 5-dimensional case we use the model  where
we have a collection of 5-D-branes wrapped on a 5-torus, and an effective
string quantized by the Polyakov prescription \maldacena\ lives inside
the 5-branes. For the 4-dimensional case we also assume,
following \KT, that the
physics is given by a central charge $c=6$ effective string. 
In this case the effective string arises at a triple intersection of
the 5-branes in M-theory \KT, so that the geometrical picture is
more complicated.

By studying the minimally coupled massless scalar in the background
of a black hole corresponding to the D-brane or M-brane bound state, we
derive the absorption cross-section using the methods of 
semi-classical General Relativity. This gives us explicit formulae
for its frequency dependence. We define the left and right moving temperatures
which, in the effective string interpretation, characterize the
thermal distributions of the left and right movers.
If $T_L, T_R\ll \sqrt{T_{\rm eff}}$, then the GR greybody factor
has a non-trivial dependence on $\omega/T_L$ and $\omega/T_R$ which
exactly agrees with our calculation on the effective string side.

If, however, $T_{L,R}$ are of the order of the effective string scale,
then the situation is more complicated.
At leading order in the energy of the incident quantum we recover the result
that the cross section for absorption equals the area of the
horizon. But the next correction in energy yields a difference between
the classical greybody factor and what the 
simplest effective string calculation, that carried out in \hrs, gives.
This suggests that in this domain of parameters 
some new effects, such as the splitting and joining of effective strings,
should be taken into account. 
It is curious that if we set a certain
parameter $\beta$ in the classical greybody factor to zero, then the
agreement with the effective string greybody factor found from
the results of \hrs\ is restored.
It is hoped that
the nature of the deviation between the classical case and the 
simplest string model will lead to an improved
understanding of the excitation spectrum
of the 5-brane - string bound state system.

We also extend our
calculations to the case of charged scalar emission, and to 4-dimensional
black holes, where all the conclusions are qualitatively similar to
the $D=5$ case. 

The plan of the paper is as follows. In section 2 we write down the wave
equation for massless scalars in the background of the 5-dimensional black
hole and compute the classical absorption cross section in the domain of 
parameters 
mentioned above. In section 3 we compute the cross section expected from
the Polyakov quantization of an effective string. 
Section 4 extends the
results to the charged case, and section 5 to the 4-dimensional case.
Section 6 compares the classical and string greybody factors.
Section 7 is a discussion, with some conjectures.

\newsec{Classical absorption by $D=5$ black holes with one large charge}

We would like to study minimally coupled massless scalars moving
in the background of a $D=5$ black hole with 3 $U(1)$ charges.
The metric of such a black hole is \refs{\ATT,\cm,\CYY,\hms}
  \eqn\MetAgain{
   d s_{5}^2 = -f^{-2/3} h d t^2 + 
                 f^{1/3} 
                  \left( {h\inv} d r^2 + r^2 d \Omega_{3}^2 \right)
 \ , }
where
$$ h(r)= \left (1 - {r_0^2\over r^2}\right )\ , \qquad 
f(r)= \left (1 + {r_5^2\over r^2}\right )
\left (1 + {r_1^2\over r^2}\right ) \left (1 + {r_n^2\over r^2}\right )
\ .$$
The $l$-th partial wave of
a minimally coupled massless scalar satisfies the radial equation,
\eqn\ExactNonExtreme{
   {h\over r^3} {d\over dr}\left 
( h r^3 {d R_l\over dr} \right) + \left [ f\omega^2 -h {l(l+2)\over r^2}    
\right] R_l(r) = 0 \ .
  }
In this paper we are concerned with the following 
range of parameters,\foot{The solution of the higher partial wave 
equations in the regime $r_0, r_n \ll r_1, r_5$ was found by
Maldacena and Strominger (private communication).}
\eqn\range{ r_0, r_1, r_n \ll r_5 \ . }
It is convenient to introduce hyperbolic angles 
$\sigma_1$ and $\sigma_n$ defined by
$$ r_n^2 = r_0^2 \sinh^2 \sigma_n \ , \qquad
r_1^2 = r_0^2 \sinh^2 \sigma_1 \ .
$$
In order to find approximate solutions of \ExactNonExtreme\ we will
match the solutions in region I, where
$ r\ll r_5$, to solutions in region II, where
$r \gg r_0, r_1, r_n$. Because of the condition \range, the inner 
and outer regions have a large overlap, so that a reliable matching is
possible.

In region I, \ExactNonExtreme\ simplifies to
\eqn\regionone{\left[ \left( h r^3 \partial_r \right)^2 +
(\omega r_5)^2 (r^2+ r_1^2) (r^2 +r_n^2)      
- h l(l+2) r^4\right] R_l(r) = 0 \ .
  }
In terms of the variable $z=h(r)$, this becomes
\eqn\one{z{d\over dz}z{dR\over dz}+
[D+{C\over (1-z)}+{E\over (1-z)^2}]R_l=0 \ ,}
where
\eqn\parameters{
\eqalign{& D={1\over 4} (\omega r_5)^2 {r_1^2 r_n^2\over r_0^4}=
{1\over 4} (\omega r_5)^2 \sinh^2 \sigma_1 \sinh^2 \sigma_n \ ,
\cr
&C= {1\over 4}\left [ (\omega r_5)^2 {r_1^2+ r_n^2\over r_0^2}
+ l(l+2) \right ]= 
{1\over 4}\left [ (\omega r_5)^2 ( \sinh^2 \sigma_1 + \sinh^2 \sigma_n )
+ l(l+2) \right ]\ ,
\cr
& E= {1\over 4} \left [(\omega r_5)^2- l(l+2)\right ] \ . \cr}
}
Remarkably, \one\ may be reduced to a hypergeometric equation by
a substitution of the form
\eqn\two{R_l=z^\alpha(1-z)^\beta F(z) \ .}
After some algebra we find that, if $\alpha$ and $\beta$
satisfy
\eqn\seven{E+\beta(\beta-1)=0\ , \qquad \alpha^2+D+C+E=0\ ,}
then the equation for $F(z)$ becomes
\eqn\eleven{z(1-z){d^2 F\over dz^2}+
[(2\alpha+1) (1-z)-2\beta z] {d F\over dz}
-[(\alpha+\beta)^2+D] F=0}
which is the hypergeometric equation!
In general, the solution to
\eqn\genhyper{z(1-z){d^2 F\over dz^2}+
[C- (1+A+B) z] {d F\over dz}
-AB F=0\ ,}
which satisfies $F(0)=1$, is the hypergeometric function $F(A,B;C;z)$.
Thus, the solution in the inner region is
\eqn\inner{R_I= z^\alpha (1-z)^\beta F(\alpha+\beta+ i\sqrt D,
\alpha+\beta- i\sqrt D; 1+ 2\alpha; z)
\ .}
Since we are interested in an incoming wave at the horizon, we choose
$$\alpha = -i \sqrt{ D+C+E} =-i {\omega r_5\over 2} \cosh \sigma_1 
\cosh \sigma_n
\ .$$
We will chose $\beta$ to be the smaller of the two roots of the quadratic
equation for $\omega r_5 < l+1$,
$$2\beta= 1- \sqrt{ 1- 4E}= 1- \sqrt {(l+1)^2-(\omega r_5)^2}
\ .$$
First we present the calculation in the regime 
$\omega r_5 < l+1$, which
is of primary interest to us. 
The modifications to $\omega r_5 >l+1$, where
$\beta$ becomes complex, will be discussed in the Appendix.

Using the asymptotics of the hypergeometric functions for 
$z\rightarrow 1$, we find that,
for large $r$, 
\eqn\asymptotic{R_I\rightarrow
\left ({r_0\over r}\right )^{2\beta} {\Gamma (1+2\alpha) \Gamma(1-2\beta)
\over \Gamma (1+ \alpha- \beta -i \sqrt D) 
\Gamma (1+ \alpha- \beta +i \sqrt D) }
}

In the outer region (region II), \ExactNonExtreme\ simplifies to
\eqn\regiontwo{ r^{-3} {d\over dr} r^3 {dR\over dr}+
\omega^2   + {(\omega r_5)^2- l(l+2)\over r^2} R_l(r) =0 \ ,
}     
which is easily solved in terms of the Bessel functions.
The dominant solution, which matches to the asymptotic form in region
I, is
$$R_{II}=2 A\rho^{-1} J_{1-2\beta} (\rho)\ , \qquad \rho= \omega r
\ .$$
For small $r$ this approaches
$${A\over \Gamma (2-2\beta)} \left ({2\over \omega r}\right )^{2\beta}
\ .$$
Matching $R_{II}$ to $R_I$,
we find that
\eqn\ppthree{ A= (\omega r_0/2)^{2\beta} (1-2\beta)
{\Gamma (1+2\alpha) \Gamma^2(1-2\beta)
\over \Gamma (1+ \alpha- \beta -i \sqrt D) 
\Gamma (1+ \alpha- \beta +i \sqrt D) }
}

The absorption cross-section may 
now be obtained using the method of fluxes.
 The flux per unit solid angle is 
  \eqn\Flux{
   {\cal F} =  \df{1}{2\i} (R^* h r^3 \partial_r R - {\rm c.c.}) \ .
  }
 The absorption probability is the ratio of the incoming flux at
the horizon to the incoming flux at infinity,
  \eqn\AbsProb{
   P= {{\cal F}_{\rm h} \ov  {\cal F}_\infty^{\rm incoming}} 
    = {\pi \omega^3\over 2} r_5 r_0^2 \cosh \sigma_1 \cosh \sigma_n
\ |A|^{-2} \ .
  }
While this expression is quite complicated in general, it simplifies
for the energy low enough that
$\omega r_5 \ll l+1$. Now $\beta\approx -l/2$ and
$A$ may be expressed in the following form,
\eqn\norm{ A=(1+l) (l!)^2 (\omega r_0/2)^{-l}
 {\Gamma \left (1- i {\omega\over 2\pi T_H}\right )\over
\Gamma \left (1+{l\over 2}- i {\omega\over 4\pi T_L} \right ) 
\Gamma \left (1+{l\over 2}- i {\omega\over 4\pi T_R} \right ) }
}
where the left and right temperatures are
\eqn\temps{ {1\over T_L}= 2\pi r_5 \cosh (\sigma_1 - \sigma_n) \ ,
\qquad {1\over T_R}= 2\pi r_5 \cosh (\sigma_1+ \sigma_n) \ .
}
As we show later, these are 
precisely the temperatures on the fractionated D-string moving within
the 5-branes.
Since the inverse Hawking temperature is
$$ {1\over T_H} = 2\pi r_5 \cosh \sigma_1 \cosh \sigma_n
$$
we have the relation \mast
$$ {2\over T_H}= {1\over T_L}+ {1\over T_R}
\ .$$

Let us note that the absorption probability for a partial wave
with $l>0$ is suppressed compared to the s-wave by a power of the
small quantity $\omega r_0$, as long as $\omega r_5 <l+1$.
Thus, for $\omega r_5< 2$
the s-wave dominates the absorption.\foot{
For $\omega r_5\ge 2$ the $l=1$ partial wave is no longer suppressed
compared to $l=0$. It is interesting that $\omega =2/r_5$ is precisely
the energy required to create the first massive state of the effective
string, whose rest energy is $\sqrt{8\pi T_{\rm eff}}$.
In general, the $l$-th partial wave becomes important for
$\omega r_5\ge l+1$.} 
Now the absorption cross-section is related to the s-wave absorption
probability by
  \eqn\SigmaAbs{
   \sigma_{\rm abs} = \df{4 \pi}{\omega^3} P_{l=0}=
2\pi^2 r_5 r_0^2 \cosh \sigma \cosh \delta
\ |A_{l=0}|^{-2}\ . }
For $\omega r_5 \ll 1$ we  find
\eqn\ppfour{ \sigma_{\rm abs}= 2\pi^2 r_5 r_0^2 \cosh \sigma_1 \cosh \sigma_n
{\omega\over 2(T_L+ T_R)} 
{e^{\omega\over T_H} - 1\over \left (e^{\omega\over 2 T_L} - 1\right )
  \left (e^{\omega\over 2 T_R} - 1 \right ) } }
which can be written as
\eqn\absorb{ \sigma_{\rm abs}= \pi^3 r_5^2 r_0^2 (1+ \sinh^2 \sigma_1
+\sinh^2 \sigma_n) 
{\omega \left ( e^{\omega\over T_H} - 1\right )
\over \left (e^{\omega\over 2 T_L} - 1\right )
  \left (e^{\omega\over 2 T_R} - 1 \right ) } \ .}
This greybody factor has  form similar to that found
in \mast ; we discuss the relations between what we
have and the results in \mast\ in section 5. 

  From detailed balance it follows that the differential rate of
Hawking emission is
\eqn\emit{ d \Gamma = \pi^3 r_5^2 r_0^2 (1+ \sinh^2 \sigma_1 
+\sinh^2 \sigma_n) 
{\omega\over \left (e^{\omega\over 2 T_L} - 1\right )
  \left (e^{\omega\over 2 T_R} - 1 \right ) } {d^4 k \over (2\pi)^4}
}
In the next section, we compare \SigmaAbs\ with the 
analysis of absorption by highly excited strings.

\newsec{The effective string calculation}

The effective string is taken to have a mass
spectrum that resembles that of the Polyakov quantized elementary string,
though we take $c=6$ rather than $c=12$ 
in the light cone gauge quantization \dvv.
While this is not likely to be an exact  quantization of the
string, we take it as an 
approximate description for large excitation levels.
Thus, we assume that the mass levels are given by
\eqn\mass{ m^2=
\left (2 \pi R n_w T_{\rm eff}+{n_p\over R}\right )^2 
+8\pi T_{\rm eff} N_R=
\left (2 \pi R n_w T_{\rm eff}-{n_p\over R}\right )^2 +8\pi T_{\rm eff} N_L
}

In the effective string calculation 
the absorption of the scalar is described by a 3-string vertex:
the initial string absorbs a massless string state
and yields a  more energetic string. 
Let the incoming scalar have energy $\omega$. Since it does not carry any 
charge, the numbers $n_w, n_p$ are not altered by the absorption of the
scalar. Thus we have
\eqn\fourone{2m\delta m= 8\pi T_{\rm eff} \delta N_R
= 8\pi T_{\rm eff} \delta N_L}
 The scalar
contributes one left oscillator $\alpha_i$ and
one right oscillator $\tilde\alpha_j$, with
\eqn\fourthree{i=j=\delta N_R=\delta N_L=
{m\over 4\pi T_{\rm eff}}\omega }

We are interested in averaging the absorption rate over all
initial string states of a given mass. Methods for performing
such a calculation were developed in \hrs.
The initial state of the excited string is assumed to
be populated by oscillators in
a thermal form \hrs , and these contribute Bose enhancement
factors to the calculation of the absorption cross-section.
Following the methods of \hrs, we find that the absorption cross-section,
which is the absorption rate minus the emission rate, is
\eqn\stringrate{ \sigma_{\rm abs}={T_{\rm eff} \over m \omega}
(2\pi \kappa_5 \delta N_L)^2 
{e^{\beta_L^* \delta N_L
+\beta_R^* \delta N_R}-1\over \left (e^{\beta_L^* \delta N_L} -1\right) 
\left (e^{\beta_R^* \delta N_R} -1\right )} 
}
where
\eqn\fourfour{\beta^*_{R,L}= {\pi\over \sqrt{N_{R,L} }}\ ,}
and $\kappa_5$ is the 5-dimensional gravitational constant.
For the right movers, the factor in the exponent is
\eqn\fourfive{{\beta_R^*\delta N_R }=
 {m\omega\over 4 T_{\rm eff}\sqrt{N_R}} \ .}
To compare with \emit, we have to identify this with 
${\omega\over 2 T_R} $.  Thus,
$$ T_R= 2 T_{\rm eff} {\sqrt{N_R}\over m}
\ .$$
Following Maldacena, we will identify the effective string
tension as the D-string tension divided by the number of
5-branes, $n_5$,
$$ T_{\rm eff}= {1\over 2\pi g n_5} = {1\over 2\pi r_5^2}
$$
where we have set $\alpha'=1$.
Using
\eqn\foureight{m={RVr_0^2\over 2g^2}
(\cosh(2\sigma_1)+\cosh(2\sigma_n))= {2\pi^2 r_0^2\over \kappa_5^2}
(1+ \sinh^2 \sigma_1 +\sinh^2 \sigma_n)}
\eqn\fournine{n_w={ n_5 Vr_0^2\over 2g}\sinh(2\sigma_1)}
\eqn\fourten{n_p={R^2 Vr_0^2\over 2g^2}\sinh(2\sigma_n)}
we find
\eqn\foureleven{[m^2-(2 \pi R
n_w T_{\rm eff}+{n_p\over R})^2]^{1/2}
={VRr_0^2\over 2g^2}2\cosh(\sigma_1+\sigma_n)\ . }
Since
\eqn\fourseven{{\sqrt{N_R}\over m}=({1\over 8\pi T_{\rm eff}})^{1/2}
[m^2-(2\pi R n_w T_{\rm eff}+{n_p\over R})^2]^{1/2}{1\over m}}
and
\eqn\fourtwelve{(\cosh(2\sigma_1)+\cosh(2\sigma_n))
=2\cosh(\sigma_1+\sigma_n)\cosh(\sigma_1-\sigma_n)}
we finally have
\eqn\fourthir{T_R= {1\over 2\pi r_5 \cosh(\sigma_1+\sigma_n)}\ .}
Performing an analogous comparison for the left-movers,
we  find 
\eqn\fourfourt{T_L= {1\over 2\pi r_5 \cosh(\sigma_1-\sigma_n)}\ .}
Thus, $T_L$ and $T_R$ are in agreement with the temperatures 
\temps\ found in the
GR greybody factor in the limit $\omega r_5 \ll 1$.
In this limit, the cross section \stringrate\ is in complete
agreement with the GR result, \absorb.

\newsec{Scalars carrying Kaluza-Klein charge}

The case of charged scalars propagating in the 
non-extremal $D=5$ black hole background 
was studied in \mast. It was found that the wave equation for
a scalar of energy $k_0$ and charge $k_5$ is obtained from that 
for a neutral scalar of energy $\omega$ by the following
replacement of parameters:
$$ \omega \rightarrow \omega'\ , \qquad \sigma_n\rightarrow
\sigma'_n\ ,
$$
where 
$$\omega'=\sqrt{ k_0^2 - k_5^2}\ , \qquad
e^{\sigma'_n} = e^{\sigma_n} {k_0- k_5\over \omega'}
$$

We wish to see if this relation also exists between the
cases of charged and neutral absorption in the effective string calculation.
We let the incoming scalar carry the $U(1)$ charge corresponding
to the momentum component $k_5$, which is
along the direction in which the effective string
carries momentum amd winding. The absorption
of the charged scalar is still given by the 3-string vertex,
 but now $\delta N_L$ is no longer equal
to $\delta N_R$. In varying the relations
\mass, we have to include the variation $\delta n_p$, whose relation
to the absorbed charge is
$$ k_5 = {\delta n_p\over R}
$$
while $ k_0= \delta m$.
Thus, we find
\eqn\chargev{2m k_0= 8\pi T_{\rm eff} \delta N_R+
2 k_5 \left (2 \pi R n_w T_{\rm eff}+{n_p\over R}\right )}
\eqn\chargevo{2m k_0= 8\pi T_{\rm eff} \delta N_L
-2 k_5 \left (2 \pi R n_w T_{\rm eff}-{n_p\over R}\right )}
Solving for $\delta N_{L,R}$, and using
\fourfour, we arrive at
$$ \beta_L^*\delta N_L = \pi r_0 \left ( k_0 \cosh (\sigma_1 -\sigma_n)
+ k_5 \sinh (\sigma_1 -\sigma_n)\right )
$$
$$ \beta_R^*\delta N_R = \pi r_0 \left ( k_0 \cosh (\sigma_1 +\sigma_n)
- k_5 \sinh (\sigma_1 +\sigma_n)\right )$$
Now it is not hard to 
see that
$$\beta_R^*\delta N_R = {\omega'\over 2 T'_R}\ ,
\qquad \beta_L^*\delta N_L = {\omega'\over 2 T'_L}
$$

We then find that the effective string
absorption cross section for charged
scalars can be written as
\eqn\chargeabs{ \sigma_{\rm abs} = \pi^3 r_5^2 r_0^2 (1+ \sinh^2 \sigma_1 
+\sinh^2 \sigma'_n) 
{\omega' \left (e^{\omega'\over T'_H} - 1\right )
\over \left (e^{\omega'\over 2 T'_L} - 1\right )
  \left (e^{\omega'\over 2 T'_R} - 1 \right ) } 
}
where 
\eqn\temps{ {1\over T'_L}= 2\pi r_5 \cosh (\sigma_1 - \sigma'_n) \ ,
\qquad {1\over T'_R}= 2\pi r_5 \cosh (\sigma_1+ \sigma'_n) \ .
}
and
\eqn\ppone{{1\over T'_H}={1\over 2 T'_L}+{1\over 2 T'_R}\ .}
This is in agreement with the GR absorption cross section in the limit
$\omega r_5 \ll 1$.

\newsec{ $D=4$ black holes with two large charges}

In this section
we turn to minimally coupled massless scalars moving
in the background of a $D=4$ black hole with 4 $U(1)$ charges.
3 of the charges may be taken to be the same as those in
the $D=5$ case: namely the D-5-brane charge, the D-1-brane charge
along one of the compact directions, and the momentum charge along
this same direction. The fourth charge can be taken to arise from
Kaluza-Klein monopoles \refs{\myers,\mst}. A different picture, which 
arises upon embedding the $D=4$ black holes into M-theory,
is to view them as triply intersecting 5-branes, wrapped over $T^7$
\KT. The radii $r_1, r_2, r_3$ are determined by
the numbers of 5-branes positioned in $(12345)$, $(12367)$
and $(14567)$ planes respectively.
We are concerned here with the situation where $r_2$ and
$r_3$ are large compared to the other radii. As shown in \HK,
in this regime the Bekenstein-Hawking
entropy coincides with the entropy of a gas
of effective strings with central charge $c=6$ and tension
\eqn\newtension{
T_{\rm eff} ={1\over 8\pi r_2 r_3}
\ .}
These strings originate from
the $(12345)$ 5-branes wrapped over the torus.

First we compute the classical absorption.
The metric of such a black hole is \refs{\CY,\CTT,\CT}
  \eqn\MetAgain{
   d s_{5}^2 = -f^{-1/2} h d t^2 + 
                 f^{1/2} 
                  \left( {h\inv} d r^2 + r^2 d \Omega_{2}^2 \right)
 \ , }
where
$$ h(r)= \left (1 - {r_0\over r}\right )\ , \qquad 
f(r)= \left (1 + {r_1\over r}\right )\left (1 + {r_2\over r}\right )
\left (1 + {r_3\over r}\right ) \left (1 + {r_n\over r}\right )
$$
The $l$-th partial wave of
a minimally coupled massless scalar satisfies the radial equation,
\eqn\FourNonExtreme{
   {h\over r^2} {d\over dr}\left 
( h r^2 {d R\over dr} \right) + \left [ f\omega^2 -h {l(l+1)\over r^2}    
\right] R_l(r) = 0 \ .
  }
We are concerned with the range of parameters
\eqn\range{ r_0, r_1, r_n \ll r_2, r_3 \ . }
It is convenient to introduce hyperbolic angles 
$\sigma_1$ and $\sigma_n$ defined by
$$ r_n = r_0 \sinh^2 \sigma_n \ , \qquad
r_1 = r_0 \sinh^2 \sigma_1 \ .
$$
In order to find approximate solutions of \ExactNonExtreme\ we will
again match the solutions in the inner and outer regions.

In region I, where $r\ll r_2, r_3$, it is convenient to use
the variable $z=h(r)$. The equation assumes the form
\one, with the parameters now defined by
\eqn\Fourparameters{
\eqalign{& D=\omega^2 r_2 r_3 
\sinh^2 \sigma_1 \sinh^2 \sigma_n \ ,
\cr
&C= 
\omega^2 r_2 r_3 ( \sinh^2 \sigma_1 + \sinh^2 \sigma_n )
+ l(l+1) \ ,
\cr
& E= \omega^2 r_2 r_3 - l(l+1) \ . \cr}
}
Thus, the solution in the inner region is given by \inner, with
$$\alpha = -i \omega \sqrt{r_2 r_3}\cosh \sigma_1 
\cosh \sigma_n
$$
$$2\beta= 1- \sqrt {(2l+1)^2-4 \omega^2 r_2r_3}
\ .$$
We will present the calculation in the regime 
$2\omega \sqrt{r_2r_3} < 2l+1$. 

In the outer region ($r\gg r_0, r_1, r_n$), \FourNonExtreme\ simplifies to
\eqn\regiontwo{ \rho^{-2} {d\over d\rho} \rho^2 
{dR_l\over d\rho}+ \left [1+ {\omega (r_2 + r_3)\over \rho}
  + {\omega^2 r_2 r_3 - l(l+1)\over \rho^2}\right ]  R_l =0 \ ,
}     
where we have defined $\rho=\omega r$.
This equation can be   solved in terms of the Coulomb functions.
The dominant solution, which matches to the asymptotic form in region
I, is
$$R_{II}= A\r^{-1}F_{-\beta}(\rho)  
\ .$$
For small $\rho$ 
$$ F_{-\beta}\approx C_{-\beta}(\eta)\rho^{1-\beta}
$$
where
\eqn\pone{\eta=-{1\over 2}(r_2+r_3)\omega}
\eqn\ptwo{C_L(\eta)={2^Le^{-\pi L/2}|\Gamma(L+1+i\eta)|\over
\Gamma(2L+2)}}

Matching $R_{II}$ to the asymptotic form of $R_I$, given in \asymptotic,
we find that
\eqn\pthree{A={(\omega r_0)^{\beta}\over C_{-\beta}(\eta)}
 {\Gamma (1+2\alpha) \Gamma(1-2\beta)
\over \Gamma (1+ \alpha- \beta -i \sqrt D) 
\Gamma (1+ \alpha- \beta +i \sqrt D) }
 }
For $2\omega \sqrt{r_2 r_3}<3$, the s-wave dominates the absorption, and we
find
\eqn\swave{\sigma_{\rm abs}=4\pi\sqrt{r_2r_3}r_0\cosh\sigma_1\cosh\sigma_n
|A_{l=0}|^{-2}\ .}

Let us now consider the effective string calculation, assuming that
the string again has $c=6$. 
The relations \fourone\ - \fourthree\ hold again.
Using \newtension, and the results in \refs{\KT,\HK,\ktI}, we find that
the relations \foureight\ - \fourten\ are replaced by
\eqn\foureightpp{m={\pi r_0\over \kappa_4^2}
(\cosh(2\sigma_1)+\cosh(2\sigma_n)) }
\eqn\fourninepp{2\pi R n_w T_{\rm eff}={\pi r_0\over \kappa_4^2}
\sinh(2\sigma_1)}
\eqn\fourtenpp{{n_p\over R}={\pi r_0\over \kappa_4^2} \sinh(2\sigma_n)}
where $\kappa_4$ is the gravitational constant in $D=4$.

We then find
\eqn\fourthirpp{T_R= {1\over 4\pi \sqrt{r_2r_3} \cosh(\sigma_1+\sigma_n)}, ~~~
T_L= {1\over 4\pi \sqrt{r_2r_3} \cosh(\sigma_1-\sigma_n)}}
  From \stringrate\ it follows that the absorption 
cross section predicted by the
simplest effective string analysis is
\eqn\psix{\sigma_{\rm abs}=2\pi\sqrt{r_2r_3}\omega r_0
[\cosh(2\sigma_1)+\cosh(2\sigma_n)]
{\omega \left ( e^{\omega\over T_H} - 1\right )
\over \left (e^{\omega\over 2 T_L} - 1\right )
  \left (e^{\omega\over 2 T_R} - 1 \right ) } }
This is in agreement with the 
$\omega \sqrt{r_2 r_3} \ll 1$ limit of the
classical cross section \swave.

In the limit $\omega\rightarrow 0$ we recover
\eqn\pfive{\sigma_{\rm abs}=4\pi\sqrt{r_2r_3}r_0\cosh\sigma_1\cosh\sigma_n}
which is the area of the horizon.

\newsec{Comparing  classical and `effective string'  greybody factors}

We now wish to carefully
compare the greybody factors obtained from the classical
calculation and from the effective string calculation. We carry out
the discussion below for the 5-dimensional case. The 4-dimensional
case is essentially similar.

First we consider the limit where $r_1\gg r_n$, which implies
$\sigma_1\gg \sigma_n$.
We can now further consider the limit $\omega r_5\ll 1$. Then
\eqn\pptwo{\beta\approx {(\omega r_5)^2\over 4}}
and we can thus ignore $\beta$ in the expression \ppthree\ for the
amplitude $A$. Then the classical greybody
factor reduces to \absorb , which equals the result \stringrate\ from the
effective string calculation.

Note that whenever $\sigma_1\gg \sigma_n$ we have
\eqn\eighttwo{\cosh(\sigma_1\pm\sigma_n)\approx {1\over 2}e^{\sigma_1}
e^{\pm\sigma_n}}
so that we can approximate the $\cosh$ functions in the classical
greybody factor by exponential. Indeed, the error in the replacement of
the $\cosh$ function by the exponential arises in 
$\sigma_{\rm abs}$ in the form
\eqn\ppfive{(\omega r_5)^2\cosh^2(\sigma_1\pm\sigma_n)
- (\omega r_5)^2{1\over 4} e^{\sigma_1} e^{\pm\sigma_n}
\approx {1\over 2}(\omega r_5)^2 }
so that if we wish to ignore $\beta$ then we must also ignore
the difference between the $\cosh$ function and the exponential.

In this limit $r_1\gg r_n$ we then find that our result reduces to
the result for the greybody factor in \mast , where the parameters
were taken to satisfy $r_0, r_n \ll r_1, r_5$. We can also compute from our
results the case $r_1\ll r_n$, which has a similar treatment. The
greybody factors resulting here again agree with the effective string
calculation, and are related to the case $r_n\ll r_1$ by duality.

Now we address the case $r_1\sim r_n$, which implies
$\cosh(\sigma_1\pm\sigma_n)\sim 1$.  First consider the limit
$\omega\rightarrow 0$. Then $\beta\rightarrow 0$, and both
the classical cross section \SigmaAbs\ and the effective
string cross section \stringrate\
 reduce to the area ${\cal A}$ of the horizon. Now we are interested
in the lowest corrections, which
are terms of order $\omega^2$. These arise from
three different sources:

(a)\quad The corrections that come from the 
$\omega/(4 \pi T_{L,R})$ terms contained 
both in the classical result and in the effective
string result - these are
of the form
\eqn\eightone{\sigma_{\rm abs}\rightarrow {\cal A}[1+C_1(\omega r_5)^2
\cosh^2(\sigma_1\pm\sigma_n)]}
where $C_1$ is a constant of order unity.

(b)\quad  The corrections that arise in the classical result (but
are not present in the effective string result) due to the
term $\beta$ in the $\Gamma$ functions in \ppthree . These are of the form
\eqn\eightthree{\sigma_{\rm abs}\rightarrow {\cal A}[1+C_2(\omega r_5)^2]}
where $C_2$ is of order unity.

(c)\quad The correction that comes from the first factor in
\ppthree\ (which is not present in the string result):
\eqn\seventhree{\sigma_{\rm abs}\rightarrow
{\cal A}({\omega r_0\over 2})^{-4\beta}\approx {\cal A}e^{-(\omega r_5)^2
\log ({\omega r_0\over 2})}\approx {\cal A}[1-(\omega r_5)^2
\log ({\omega r_0\over 2})]}

Note that
\eqn\eightfive{\omega r_0=(\omega r_5)(r_0/r_5)\ll 1}
so that
\eqn\eightsix{|\log ({\omega r_0\over 2})|\gg 1}

Thus we see that the leading correction comes from the last source (c).
Note that we are probing a domain where $\omega r_5= 
\omega\sqrt{\alpha'_{\rm eff}}$ 
is no longer infinitesimal, so that the dynamics on the effective string 
scale comes into play.  (When 
$r_1$ and $r_n$ are both comparable to $r_0$, the temperatures
on the effective string are sufficiently high that nontrivial greybody
factors arise only for  $\omega r_5$ non-infinitesimal.)

\newsec{Discussion}

We have computed the greybody factors in the classical geometry and in
the effective string model. 
Quantizing the string by the Polyakov method incorporates the  duality
symmetry expected of the greybody factors,
 which interchanges the string winding and the momentum charges.
This symmetry interchanges the 1-brane charge with the momentum charge
and is part of the S-duality, but from the point of view of the
effective string it is T-duality.
Thus the string calculation covers the domains where the winding dominates
over momentum, where the momentum dominates over winding, and
where they are comparable. The left and right temperatures
of the string have the form \temps, which 
reduce to the results of \mast\ in the  limit where the winding charge is
much larger than the momentum charge, and its T-dual result when
the momentum charge is much larger than the winding charge.
In these domains, the left and right temperatures are much smaller
than $1/r_5$, and for scalar energy comparable to $T_L, T_R$
the GR greybody factor is reproduced by an effective string
calculation.

Our GR calculation is also valid when $T_L$ and $T_R$ are comparable
to $1/r_5$. Now the greybody factor contains no energy scale
$\ll r_5$.
While there is agreement, as expected, at the
leading order in energy (the cross section is the area of the horizon in
both cases), there are deviations between the two calculations at
the next order in the energy of the incident quantum. The form of the
deviations is  suggestive of the fact that some modification of the
effective string model would yield agreement with the classical
result at higher orders in the energy. One such modification that
may be necessary is to include the splitting of the effective string
because its string coupling is of order 1. 
If the parameter $\beta$ in the classical
greybody factor \ppthree\ is set to zero, we obtain the
same greybody factors that the effective string predicts. The leading
correction when $\beta$ is nonzero is given by the term \seventhree.
This term  is reminiscent of
the `world sheet cut-off'
 dependent factor that must be multiplied with the naive
expression of a vertex operator in string theory. Since we have an 
effective string
with a noncritical central charge, it may be that such a  factor survives in
final expressions for the amplitude.\foot{It was noted in
\ref\alwissato{S.P. de Alwis and K. Sato, hepth/9611189.}
that for black holes other than the supersymmetric 5 and 4
 dimensional ones considered above, an effective string
model does not hold for the absorption
cross sections beyond  leading order in the
energy.}

We observed that, even in the limit $r_5\gg r_0,r_1,r_n$, the higher
angular momentum partial waves can contribute to the classical
absorption cross section for $\omega r_5$ of order 1.
It is suggestive that the $l=1$ partial wave begins to contribute
at exactly the energy that is needed to create 
the first massive state of the effective string.

In our calculation the effective string was taken to behave as
a single multi-wound string that can absorb and emit quanta. But
at least for a string in free space (i.e. not bound to 5-branes) such
behavior is true only in certain domains of coupling-length space
\ref\dasmathurone{S. Das and S.D. Mathur, Phys. Lett. 
B375 (1996) 103, S.D. Mathur hep-th/9609053.}. If the unexcited  string
is too short, for instance, it responds to an incoming graviton
by splitting rather than moving to a higher excited state \dasmathurthree .
In the present case the effective string is likely to be strongly coupled,
as there is no small parameter that governs its coupling. Thus we
may need to take into account multi-string processes to get better
agreement with the classical greybody factors.

In this quantization of the string the excitation levels are described
by $N_L, N_R$ which cannot be associated with either winding or momentum
excitations alone, but represent some combination of the two. 
This is natural in view of 
the T-duality of the excitation spectrum. One would like to formulate all
results on the excitation of branes in such 
duality invariant ways. For example,
we know that when $n_w$ D-strings come close to each other, they are
described by an $SU(n_w)$ gauge theory. What happens when these D-strings
also carry a total of $n_p$ units of momentum along their winding direction?
If the strings are very long, and the momentum density correspondingly dilute,
we can imagine that the effective physics is still governed by $SU(n_w)$
gauge theory. But if the momentum density is high, it is more convenient to 
perform dualities that interchange the winding and momentum charges. In
the dual picture there are $n_p$ strings, and in some domain of parameters they
are described by an $SU(n_p)$ gauge theory. What then can we say of the
general description of the situation with both winding and momentum
charges present? 

In \ref\vafa{C. Vafa, \np {\bf B463} (1996) 435.}
it was argued that, in the case of $n_w$ 0-branes
on $n_p$ 4-branes which is related to our case by T-duality, 
the moduli space is $T_4\times T_4^{n_wn_p}/S^{n_wn_p}$,
where $T_4$ is the 4-torus. At the tip of the cone generated by
the symmetrization, we have a similar  structure to that which
would  arise in the
moduli space of $n_wn_p$ 0-branes, which are described by
$SU(n_pn_w)$.Thus it is possible 
that at high density of excitations
(the limit where one expects to make a black hole) the
effective description involves the gauge group $SU(n_pn_w)$, and the
groups $SU(n_w)$, $SU(n_p)$ arise only in certain limits of moduli.

Where can such a large group come from?
Note that the product $n_pn_w$ represents the number of 
fractional momentum modes 
that arise from $n_p$ units of momentum on $n_w$ strings. Equivalently,
we can T-dualize to obtain $n_w$ 0-branes joined by
$n_wn_p$ open string segments. If these momentum modes or
open strings interact with each other, then they might yield the
large gauge group suggested above.
Thus, it is possible that when the
brane charges are increased in a compact
volume, at some point the more complicated gauge symmetry arises. There are
open fundamental strings that run between the D-strings, but there can 
also be open D-strings that
connect these open fundamental strings because, by duality, a D-string
can terminate on a fundamental string.
This may lead to
a complex nested structure of gauge groups, especially at the coupling
that is strong enough to make the D-string tension comparable to
the fundamental string tension.
Note that the black hole
horizon area depends on the product 
$n_5n_pn_w$, so that it is natural to consider
this number of elementary constituents in trying to estimate the
bound state size for branes or to realize 
the idea of `holography' \ref\susshol{L. Susskind, J. Math. Phys
{\bf 36} (1995) 6377, hep-th/9409089.} in this model.

\newsec{Appendix: absorption by $D=5$ black holes 
for $\omega r_5 \ge l+1$}

For $\omega r_5\ge l+1$, the matching has to be done more carefully.
In the inner region, the large $r$ asymptotic is
$$
R_I\rightarrow
\left ({r_0\over r}\right )^{2\beta} {\Gamma (1+2\alpha) \Gamma(1-2\beta)
\over \Gamma (1+ \alpha- \beta -i \sqrt D)
\Gamma (1+ \alpha- \beta +i \sqrt D) }
$$
$$+ \left ({r_0\over r}\right )^{2 (1-\beta)}
{\Gamma (1+2\alpha) \Gamma(2\beta-1)
\over \Gamma (\alpha+ \beta -i \sqrt D)
\Gamma (\alpha+ \beta +i \sqrt D) }
$$
Now $\beta$ and $1-\beta$ are complex conjugates,
$$ \beta= {1-i x\over 2}\ , \qquad x=\sqrt{(\omega r_5)^2-(l+1)^2}
$$
In the outer region the solution is
$$R_{II}=2 \rho^{-1} (A J_{ix} (\rho)
+ B J_{-ix} (\rho) )\ , \qquad \rho= \omega r
\ .$$
Matching the two terms separately, we find
$$ A=i x (\omega r_0/2)^{1-ix}
 {\Gamma \left (1- i {\omega\over 2\pi T_H}\right )
\Gamma^2 (ix) \over
\Gamma \left ({1+ix\over 2}- i {\omega\over 4\pi T_L} \right )
\Gamma \left ({1+ix\over 2}- i {\omega\over 4\pi T_R} \right ) }
\ ,$$
$$ B=-i x (\omega r_0/2)^{1+ix}
 {\Gamma \left (1- i {\omega\over 2\pi T_H}\right )
\Gamma^2 (-ix) \over
\Gamma \left ({1-ix\over 2}- i {\omega\over 4\pi T_L} \right )
\Gamma \left ({1-ix\over 2}- i {\omega\over 4\pi T_R} \right ) }
\ .$$
The reflection coefficient for the $l$-th partial wave is
$$ {\cal R}_l = {A e^{\pi x/2} + B e^{-\pi x/2}\over
A e^{-\pi x/2} + B e^{\pi x/2} }
\ .$$

In the limit of large $\omega r_5$, it is possible to show that
the reflection probability satisfies the following bound,
$$ |{\cal R}_l|^2 \leq  4 e^{-2\pi \omega r_5}
\ .$$
Thus, for each partial wave, we find that the reflection probability
vanishes exponentially at high energies.
This agrees with the intuition that, at high energies, black holes are
good absorbers.

\newsec{Acknowledgements}

We are grateful to C.G.~Callan, S.S.~Gubser, J.~Maldacena and
A.A.~Tseytlin for useful discussions and comments.  The work of
I.R.K. was supported in part by DOE grant DE-FG02-91ER40671,
the NSF Presidential Young Investigator Award PHY-9157482, and the
James S.{} McDonnell Foundation grant No.{} 91-48. S.D.M. was
partially supported by DOE cooperative agreement no. DE-FC02-94ER40818.

\listrefs
\bye